# Impact of Synchronous Condensers on Power System Static Voltage Stability Considering Line Contingencies in the Presence of Renewable Generation

**Umair Shahzad**

Department of Electrical and Computer Engineering, University of Nebraska-Lincoln, Lincoln, NE, USA
Email: umair.shahzad@huskers.unl.edu

**ABSTRACT**

Ever-growing electrical loads are having a huge impact on the operation and stability of the power system. Moreover, the integration of renewable generation poses various challenges to the future power system, especially, regarding stability. Thus, this paper presents the impact of synchronous condensers (SCs) on static voltage stability analysis for a test transmission network, considering line contingencies, in the presence of renewable generation. The main purpose of this study was to identify critical buses in the power system when line contingencies occur. Both *(N-1)* and *(N-2)* contingencies were considered in this study. The impact of renewable generation is also assessed. To analyze the static voltage stability, the conventional power-voltage (P-V) curve method, using continuation power flow (CPF), is applied on the IEEE 14-bus test system. DIgSILENT PowerFactory software simulations are used to obtain the results. The P-V analysis accurately quantifies the critical buses for both cases, considering *(N-1)* and *(N-2)* line contingencies.

1. **INTRODUCTION**

Since the early 20th century, power system stability has been documented as a significant issue for secure system operation [1-2]. Most blackouts caused by power system instability have demonstrated the significance of this phenomenon [3-4]. Historically, transient stability has been the leading stability issue in most power networks. However, with the introduction of novel technologies and increasing load demands, several kinds of instability have come into the picture. For instance, voltage stability, frequency stability and interarea oscillations have gained importance. This has necessitated an understanding of the concept of power system stability. A lucid concept of various kinds of instability is vital for the acceptable operation of power systems. Reference [5] has largely classified power system stability into three types: rotor angle, frequency, and voltage. This is pictorially shown in Fig. 1. An account of these types of stability follows.

*Rotor angle stability* is the ability of synchronous machines in the power system to maintain synchronism when a disturbance is applied. Instability can result when the angular swing of generators leads to loss of their synchronism. Small-signal, rotor-angle stability deals with stability under small disturbances, such as minor load variations. Large-angle stability focuses on large disturbances, such as three-phase short circuit.

*Frequency stability* is the ability of a power system to maintain steady frequency after a severe system stress causes a substantial disparity between generation and load. It relies on the ability to preserve equilibrium between system supply and demand, with a minimum inadvertent loss of load. Instability can manifest itself in the shape of sustained frequency swings, which subsequently cause generating units and loads to trip. An example of this phenomenon is the forming of an under generated island with inadequate under-frequency load shedding such that frequency declines swiftly, resulting in a blackout of the island within a short span of time. Longer-term phenomena include situations in which steam turbine overspeed controls cause frequency instability.

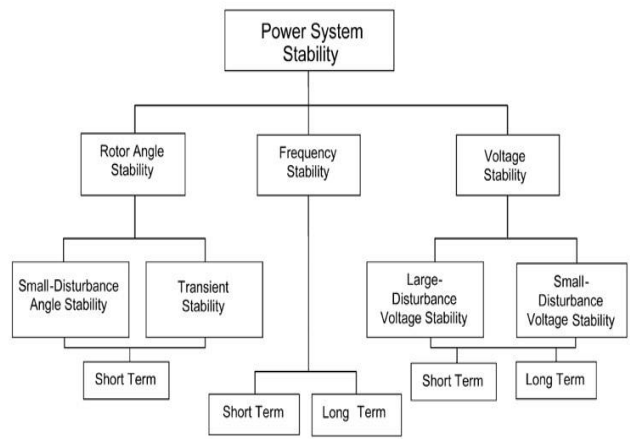

Figure 1. Classification of power system stability.

*Voltage stability* is the ability of a power system to maintain steady voltages at all buses in the system after being subjected to a disturbance from a given initial operating condition. It depends on the ability to maintain equilibrium between system demand and system generation. Instability may manifest itself in the shape of a continuing decrease or increase of voltages at some or all buses. The culprit for voltage instability is typically the loads: in response to a disturbance, power expended by the loads is apt to be reinstated, mostly due to motor slip alteration and tap-changing transformers. Reinstated loads upsurge the strain on the high voltage network by snowballing the reactive power consumption, thereby producing additional voltage drop. A deteriorating state results in voltage instability when load dynamics try to reinstate power consumption outside the transmission system capability. A term which is frequently used in this regard is voltage collapse. It is defined as the procedure by which the order of events complementing voltage instability causes a total blackout or unusually low voltages in a larger portion of the power network. Large-disturbance voltage stability is the ability of a power system to maintain steady voltages after the occurrence of large disturbances, such as three-phase faults. The inherent features of system and load determine this ability. It is essential to investigate the nonlinear response of a power system over a period of time to compute large-disturbance voltage stability. The evaluation period typically ranges from a few seconds to tens of minutes. Small-disturbance voltage stability is the system's ability to maintain steady voltages when small disturbances occur, such as gradual changes in system load. In modern power networks, factors such as congested transmission lines and increasing load demands have caused a highly stressed system. Presently, imminent voltage instability poses a grave risk to the security of these networks. There are two main techniques that are used to analyze voltage stability: static and dynamic. The former approach deals with traditional power flow solutions that are suitable for cases where precontingency and postcontingency situations are recognized for voltage stability limits. The dynamic method, however, uses highly non-linear differential equations to incorporate the dynamics of generators [6-7]. The main reason for voltage instability is the lack of reactive power. With large loading, a voltage stability issue may arise as voltages on some or all buses reach a critical point and are unable to regain their original values. Voltage instability is usually a localized process. Nevertheless, it may grow into a network-wide issue if transmission system operators (TSOs) do not take appropriate timely measures [8]. Annually, voltage instability is the reason for huge revenue losses. Voltage instability was a major cause of the 2003 Northeast U.S. blackout. Voltage instability also caused the 1987 Tokyo blackout [9].

Although, some research has been conducted on static voltage stability of transmission networks, these studies do not consider the line contingencies, which this study considered while assessing static voltage stability. Moreover, the impact of renewable generation on voltage stability is also of great significance and hence, this needs to be investigated, as well. The remainder of this paper is organized as follows. Section 2 discusses various methods for assessing static voltage stability. Section 3 discusses the computation procedure. Sections 4 and 5 describe the modeling of generation sources and relevant case studies, respectively, for the static voltage stability assessment of the IEEE 14-bus system. Section 6 discusses the results obtained. Finally, Section 7 concludes the paper with suggested future research directions.

## 2. STATIC VOLTAGE STABILITY METHODS

There are several techniques that can be used to assess static voltage stability: modal analysis, singular value decomposition, sensitivity analysis, and the power-voltage (P-V) curve method [10-11]. A brief discussion on these methods follows [12].

*A. Modal Analysis*

The technique of modal analysis is used to assess voltage stability in [13–18]. The power flow Jacobian matrix is the basis for this analysis. It has been used for classifying the critical system bus, which causes system instability. A well-defined origin can be used to study both the steady-state stability and the voltage collapse point [19]. The authors in [13] used the eigenvalue and related eigenvectors of the reduced Jacobian matrix to conduct a modal analysis. In their work, the voltage variation with the reactive power was incorporated with reference to the rank of eigenvalues. Work in [14] largely focused on the system security augmentation with reference to the voltage stability. In addition, this study determined the voltage collapse point by sensitivity and eigenvalue analysis that was applied on the Italian Electric Power Company (ENEL) transmission system. Reference [15] used the Jacobian matrix to research the geometry of the reactive power load flow. The approach was founded on the maximum power transfer point. Consequently, the margin distance to the collapse point was computed.

*B. Singular Value Decomposition*

This technique is used for estimating the reactive power compensation required to optimally distribute the resources throughout the system for attaining voltage stability. The minimum singular value of the power flow Jacobian matrix is used as a measure to compute the distance between the static voltage limit and the system operating point. Reference [20] focused on the relation between singularity of load flow in the Jacobian matrix and the singularity of dynamics of the system. The work determined that maximum loadability is determined by the singularity of load flow Jacobian. Reference [21] deals with a comprehensive research of the singular value decomposition technique. The method can approximate the point of system collapse and can identify the critical buses. Reference [22] used a weighted least square algorithm to suggest a novel static state estimation algorithm. The proposed method is based on the singular value decomposition method.

*C. Sensitivity Analysis Method*

Weak buses are the buses which have the most tendency to become unstable with respect to voltage [23]. These buses are usually identified by conducting a bus sensitivity analysis. The sensitivity analysis is used to enhance the loadability of weak buses and hence, improve the global stability of the power system. The sensitivity index for bus voltage is considered as the bus voltage changes with reference to $\frac{\Delta V_i}{\Delta Q_i}$ and $\frac{\Delta V_i}{\Delta P_i}$ changes [24-25]. Nevertheless, the sensitivity index is not adequate to classify weak buses, particularly, in an interconnected network [26-27]. Compared to other approaches, sensitivity analysis is of greater significance in the identification of critical buses in the network. It is, however, imperative to examine how altering the network situations impact the critical point. Moreover, sensitivity analysis can be used for computing the reactive power margin (MVar distance to the voltage collapse point) of the weak buses [28-29].

*D. P-V Curve Method*

Power system voltage stability focuses on the relationship between transmitted power (P) and receiving end voltage (V). A conventional way to display this relationship is through a P-V curve, which is attained using a steady-state analysis. For this analysis, P, i.e., system active power, is increased in steps; and the voltage (V) is observed at system buses. This technique uses continuation power flow (CPF). The CPF starts with a base load, and consequently, computes the maximum transfer power by increasing the load in discrete steps. Therefore, the loadability margin can be computed which represents the maximum active load at the critical buses in the power network. Consequently, curves for these buses are plotted to attain the voltage stability of a system. The relationship between bus voltage and MW transfer is nonlinear, which necessitates the full power flow solutions [30]. According to the philosophy of this method, the system bus is stable if the operating point is above the nose point, as shown in Fig. 2. However, the system bus is unstable if the operating point lies in the lower portion of the P-V curve. After the nose point, the load flow no longer converges. The distance between the operating point and the nose point is known as the stability margin at that bus. The nose point is also known as the knee point or critical point.

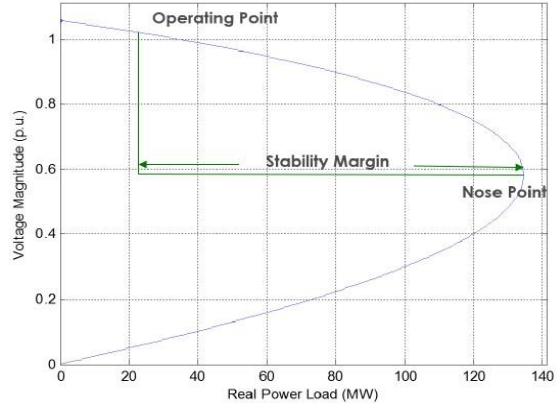

Figure 2. P-V curve method.

This paper describes a P-V analysis to identify critical buses for a test transmission network in the presence of *(N-1)* and *(N-2)* line contingencies. The novelty of the work is the consideration of both *(N-1)* and *(N-2)* line contingencies, in identifying critical buses, using static voltage stability assessment, considering the effect of synchronous condensers (SCs). The impact of integrating renewable energy generation is also studied.

## 3. COMPUTATION PROCEDURE

The IEEE 14-bus test system was used to conduct the required studies. Two cases were considered. In Case 1, SCs were disconnected from the system, whereas in Case 2, they were connected. The computation procedure for identifying critical system bus(es) for both cases is shown in Fig. 3. In the first step, all (N-1) and (N-2) transmission line contingencies were defined for the network. A P-V analysis was conducted for each (N-1) and (N-2) line contingency using DIgSILENT PowerFactory software After conducting the analysis, P-V curves were plotted for each bus. From the resulting plots, critical buses were identified. The next step was to integrate renewable generation sources, such as wind generation (squirrel cage induction generator and doubly-fed induction generator), and photovoltaic (PV) generator.

## 4. MODELING OF GENERATION SOURCES

*A. Synchronous generator*

Standard 6$^{th}$ order model is used for modeling all synchronous generators. All synchronous machines are equipped with (TGOV1) turbine governor, (IEEEX1) exciter (including PI type automatic voltage regulator), and (STAB1) power system stabilizer. The associated mathematical model for a standard 6$^{th}$ order synchronous generator is given by following equations [31].

$$T_J \frac{d\omega}{dt} = M_m - M_c - D(\omega - \omega_o) \quad (1)$$

$$\frac{d\delta}{dt} = \omega - \omega_o \quad (2)$$

$$T'_{do} \frac{dE'_q}{dt} = E_{fd} - (x_d - x'_d)I_d - E'_q \quad (3)$$

$$T''_{do} \frac{dE''_q}{dt} = -E''_q - (x'_d - x''_d)I_d + E'_q + T''_{do} \frac{dE'_q}{dt} \quad (4)$$

$$T'_{qo}\frac{dE'_d}{dt} = -E'_d + (x_q - x'_q)I_q \quad (5)$$

$$T''_{qo}\frac{dE''_d}{dt} = -E''_d - (x'_q - x''_q)I_q + E'_d + T''_{qo}\frac{dE'_d}{dt} \quad (6)$$

Where $T_J$ is generator inertia constant; $M_m$ is mechanical torque; $M_c$ is electromagnetic torque; $D$ is damping coefficient; $\delta$ is rotor angle; $\omega$ is rotor speed; $\omega_o$ is synchronous speed; $E_{fd}$ is field voltage; $E'_d$ and $E'_q$ are d and q axis components of transient electric potential; $E''_d$ and $E''_q$ are d and q axis components of sub transient electric potential; $x_d, x'_d, x''_d, x_q, x'_q, x''_q$ are d and q axis synchronous reactance, transient reactance and sub transient reactance, respectively; $T'_{do}, T'_{qo}$ are d and q axis transient time constants, and $T''_{do}, T''_{qo}$ are d and q axis sub transient time constants.

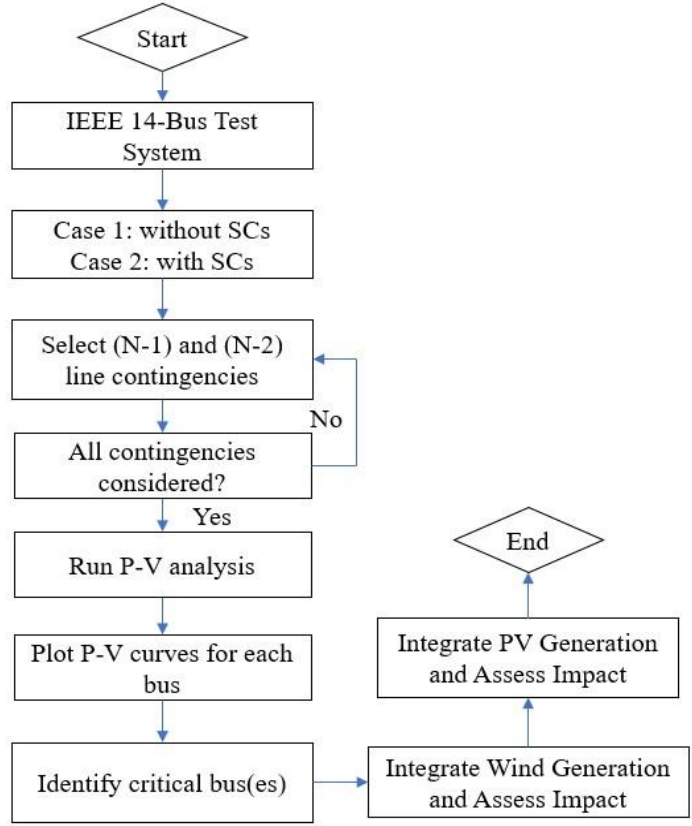

Figure 3. Computation procedure.

*B. Squirrel cage induction generator (SCIG)*

Single cage model with current displacement is issued to model squirrel cage induction wind generator. It is also equipped with compensating capacitors to limit the reactive power drawn from the grid. The mathematical equations describing SCIG model are as follows [32].

$$\frac{d\Psi_{ds}}{dt} = u_{ds} + \omega_s\Psi_{qs} + R_s i_{ds} \quad (7)$$

$$\frac{d\Psi_{qs}}{dt} = u_{qs} - \omega_s\Psi_{ds} + R_s i_{qs} \quad (8)$$

$$\frac{d\Psi_{dr}}{dt} = u_{dr} + s\omega_s\Psi_{qr} - R_r i_{dr} \quad (9)$$

$$\frac{d\Psi_{qr}}{dt} = u_{qr} - s\omega_s\Psi_{dr} - R_r i_{qr} \quad (10)$$

$$\frac{d\theta_m}{dt} = \omega_{mr} \quad (11)$$

Where all variables are in per unit. $\Psi_{ds}$, $\Psi_{qs}$ are stator flux linkages; $\Psi_{dr}$, $\Psi_{qr}$ are rotor flux linkages; $u_{ds}$, $u_{qs}$ are stator voltages; $u_{dr}$, $u_{qr}$ are rotor voltages; $i_{ds}$, $i_{qs}$ are stator currents; $i_{dr}$, $i_{qr}$ are rotor currents; $R_s$, $R_r$ are stator and rotor resistance, respectively; $\omega_s$ is synchronous electrical speed; $\theta_m$ is rotor angle; $s$ is generator slip, and $\omega_{mr}$ is mechanical rotor speed.

*C. Doubly fed induction generator (DFIG)*

A reduced 3rd order model, which neglects the stator transients, is used to represent DFIGs. The model has a structure like that proposed by WECC [33] and IEC [34]. Ignoring the stator current dynamics, the mathematical equations governing the DFIG model are as follows [35].

$$\frac{1}{\omega_b}\frac{de_d}{dt} = \frac{-1}{T_o}[e_d - (X - X')i_{qs}] + s\omega_s e_q - \omega_s \frac{L_m}{L_m + L_r} v_{qr} \quad (12)$$

$$\frac{1}{\omega_b}\frac{de_q}{dt} = \frac{-1}{T_o}[e_q - (X - X')i_{ds}] - s\omega_s e_d + \omega_s \frac{L_m}{L_m + L_r} v_{dr} \quad (13)$$

$$\frac{d\omega_r}{dt} = \frac{1}{2H_g}[K_{tw}\theta_{tw} + D_{tw}(\omega_t - \omega_r) - (e_d i_{ds} + e_q i_{qs})] \quad (14)$$

$$v_{ds} = -r_s i_{ds} + X' i_{qs} + e_d \quad (15)$$

$$v_{qs} = -r_s i_{qs} - X' i_{ds} + e_q \quad (16)$$

$$P_w = v_{ds}i_{ds} + v_{qs}i_{qs} - v_{dr}i_{dr} - v_{qr}i_{qr} \quad (17)$$

$$Q_w = v_{qs}i_{ds} - v_{ds}i_{qs} \quad (18)$$

Where $e_d$, $e_q$ are d and q components of internal voltage; $P_w$, $Q_w$ are active and reactive power of DFIG absorbed by the network; $X$, $X'$ are open-circuit and short-circuit reactance; $T_o$ is the transient open-circuit time constant; $H_g$ is generator inertia constant; $\omega_b$, $\omega_s$, $\omega_r$ are system base speed, synchronous speed and rotor speed, respectively; $s$ is generator slip; $\theta_{tw}$ is the shaft twist angle (radians); $K_{tw}$, $D_{tw}$ are the shaft stiffness and mechanical damping coefficients; $v_{ds}$, $v_{qs}$ are stator voltages; $v_{dr}$, $v_{qr}$ are rotor voltages; $i_{ds}$, $i_{qs}$ are stator currents; $i_{dr}$, $i_{qr}$ are rotor currents; $r_s$ is stator resistance; $L_m$, $L_r$ are magnetizing and rotor inductances.

*D. Photovoltaic (PV) generator*

PV generator is modelled using a constant active power approach as suggested by presently practiced standards IEEE 1547 and UL 1741 [36,37]. This means irradiation and PV cell temperature are constant during the simulation period. As transient stability is analyzed in a time window of a few seconds (5s in this research), the assumption, therefore, is rational. The used bus model is PV type, i.e., the injected active power P and the voltage magnitude V are specified. This operation mode allows the generator to support the voltage profile in the network through injecting reactive power. Voltage control is done locally i.e., reactive power output of generator is controlled to attain the specific local voltage at its terminal.

5. **CASE STUDIES AND SIMULATIONS**

As mentioned previously, IEEE-14 bus test transmission system was used to conduct the required analysis. The single line diagram for this system is shown in Fig. 4. The system consists of 14 buses (of which 11 are load buses), 2 synchronous generators, 3 SCs (connected at Buses 3, 6 and 8) and a total of 16 transmission lines. The total load was 259 MW, and the installed capacity of the generators was 272 MW [38]. For P-V curve analysis, the system active power was subjected to an augmentation of 1 MW at each step; and corresponding voltages at system buses were observed. This was first done for the system without any contingency so that the critical bus could be identified. Then, the procedure was conducted for both (N-1) and (N-2) transmission line contingencies.

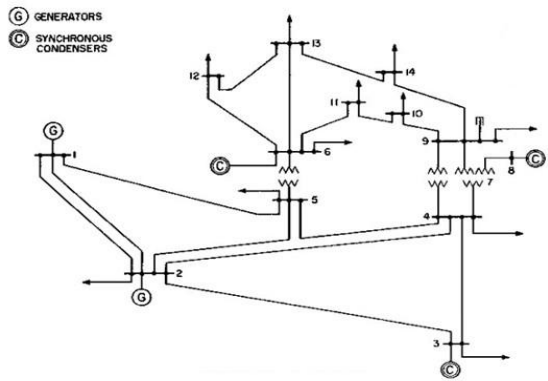

Figure 4.  IEEE 14-bus test system.

## 6.  RESULTS AND DISCUSSION

This section discusses the results of the P-V analysis, first without SCs and then including them. The SCs were connected to Buses 3, 6 and 8 in the latter analysis. For each analysis, three subcases were considered: base case, (N-1) line contingencies, and (N-2) line contingencies. An SC is essentially a DC-excited synchronous motor, whose shaft is unconnected, but it can rotate freely. It is used to regulate conditions on the power system grid rather than to convert electric energy to mechanical energy, or vice versa. Its field is controlled using a voltage regulator to either generate or absorb reactive power, as required, to adjust the voltage of the grid [39].

*Case 1: P-V analysis without considering SCs*. First, P-V curves were plotted for the base case (no contingencies). The plot for P-V curves for all of the system buses is shown in Fig. 5. It should be noted that the software stops plotting the curve as soon as the knee point is achieved; however, it is not hard to trace the curve back intuitively. A possible reason for not plotting further is that only the knee point is of interest, and there is no use in conducting further analysis once this instability point is reached. As evident, the voltage magnitude for Bus 14 at the nose point is the lowest. Therefore, Bus 14 is the critical bus.

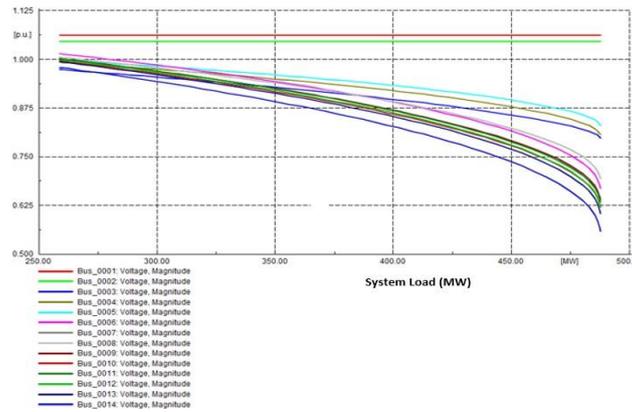

Figure 5.  P-V curves for base case (bus 14 is the critical bus).

In the next step, (N-1) line contingencies were considered. The P-V curves for each (N-1) line contingency were plotted against the voltage magnitude of system buses. For instance, P-V curves for Line 1-5 contingency are shown in Fig. 6.

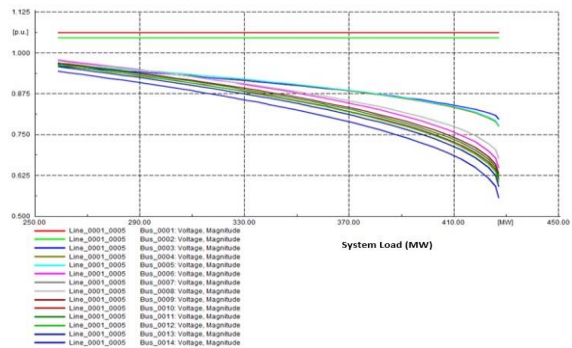

Figure 6.  P-V curves for Line 1-5 contingency (bus 14 is the critical bus).

From the curves, it is evident that Bus 14 is the critical bus. Similarly, all critical buses were identified for the remaining (N-1) line contingencies. The results are provided in Table 1.

The graphical results are shown in Fig. 7. As is evident, Bus 14 is the critical bus as it had the highest frequency for being the critical bus when (N-1) line contingences were considered. Here "frequency" is defined as the number of times a bus becomes critical.

A similar analysis was conducted for (N-2) line contingencies (amounting to 120 contingencies). The graphical results are shown in Fig. 8. Again, Bus 14 is the critical bus.

TABLE 1. IDENTIFICATION OF CRITICAL BUSES FOR (N-1) LINE CONTINGENCIES (WITHOUT SCS)

| Line Contingency Number | Line Name | Critical Bus |
|---|---|---|
| 1 | Line_0001_0002/1 | 14 |
| 2 | Line_0001_0002/2 | 14 |
| 3 | Line_0001_0005 | 14 |
| 4 | Line_0002_0003 | 3 |
| 5 | Line_0002_0004 | 14 |
| 6 | Line_0002_0005 | 14 |
| 7 | Line_0003_0004 | 14 |
| 8 | Line_0004_0005 | 14 |
| 9 | Line_0006_0011 | 14 |
| 10 | Line_0006_0012 | 12 |
| 11 | Line_0006_0013 | 13 |
| 12 | Line_0009_0010 | 10 |
| 13 | Line_0009_0014 | 14 |
| 14 | Line_0010_0011 | 14 |
| 15 | Line_0012_0013 | 14 |
| 16 | Line_0013_0014 | 14 |

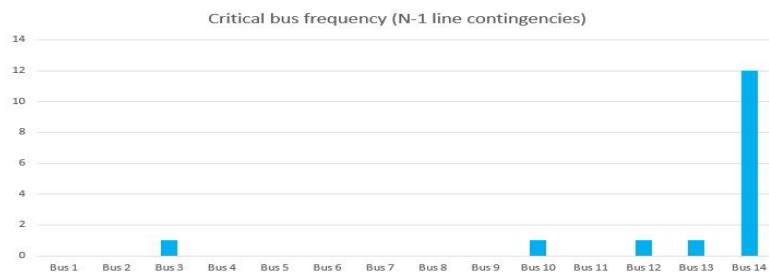

Figure 7. Bar graph showing that Bus 14 is the critical bus for (N-1) line contingencies.

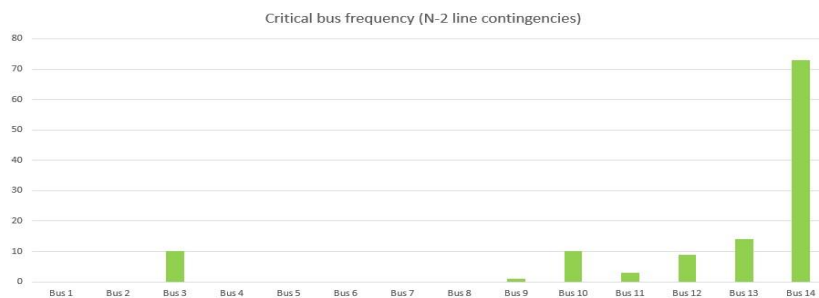

Figure 8. Bar graph showing that Bus 14 is the critical bus for (N-2) line contingencies.

*Case 2: P-V analysis considering SCs.* Now, SCs were connected to the network at Buses 3, 6 and 8. The P-V curves were plotted for the base case (no contingencies). The plot for P-V curves for all system buses is shown in Fig. 9. As evident, the voltage magnitude for Bus 5 at the nose point is the lowest. Therefore, Bus 5 is the critical bus.

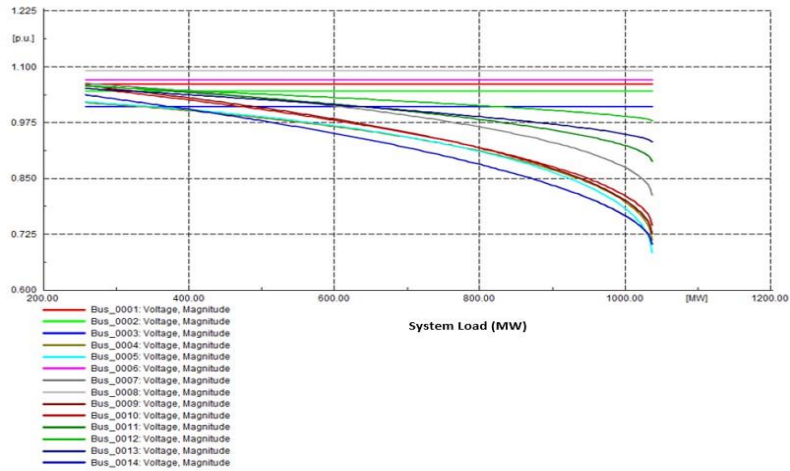

Figure 9. P-V curves for base case (bus 5 is the critical bus).

In the second case, (N-1) line contingencies were considered. The P-V curves for each (N-1) line contingency were plotted against the voltage magnitude of system buses. For instance, P-V curves for Line 1-5 contingency are shown in Fig. 10.

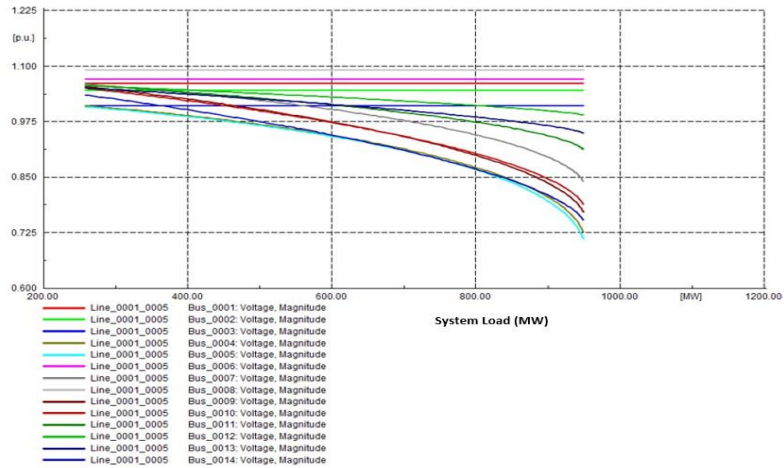

Figure 10. P-V curves for Line 1-5 contingency (bus 5 is the critical bus).

From the curves, it is evident that Bus 5 is the critical bus. Similarly, all critical buses were identified for the remaining (N-1) line contingencies. The results are provided in Table 2. The graphical results are shown in Fig. 11. As is evident, Bus 5 is the critical bus, as it has the highest frequency for being the critical bus, when (N-1) line contingences are considered. A similar analysis was conducted for (N-2) line contingencies (amounting to 120 contingencies). The results are shown graphically in Fig. 12. Again, Bus 5 is the critical bus. In conclusion, it can be said that Bus 14 is the critical bus for the IEEE 14-bus test system (without SCs) when the system operates normally (no contingency). In the presence of (N-1) and (N-2) line contingencies, the same bus is the critical bus. With the inclusion of SCs, Bus 5 is the critical bus when the system operates normally (no contingency). In the presence of (N-1) and (N-2) line contingencies, the same bus is the critical bus. Research work in [40] also validates the results we obtained for the base case. The overall summary of a P-V analysis of the IEEE 14-bus test system with regard to critical buses is shown in Table 3. Similarly, results were obtained with renewable generation integration and are shown in Tables 4-6.

Recent research works [41-43] strongly indicate that voltage stability in power systems is a growing area and an in-depth research is required to further expand its horizon, especially with the rising uncertainties in power systems [44-50].

TABLE 2. IDENTIFICATION OF CRITICAL BUSES FOR (N-1) LINE CONTINGENCIES (WITH SCs)

| Line Contingency Number | Line Number | Critical Bus |
|---|---|---|
| 1 | Line_0001_0002/1 | 5 |
| 2 | Line_0001_0002/2 | 5 |
| 3 | Line_0001_0005 | 5 |
| 4 | Line_0002_0003 | 4 |
| 5 | Line_0002_0004 | 5 |
| 6 | Line_0002_0005 | 5 |
| 7 | Line_0003_0004 | 5 |
| 8 | Line_0004_0005 | 5 |
| 9 | Line_0006_0011 | 11 |
| 10 | Line_0006_0012 | 5 |
| 11 | Line_0006_0013 | 13 |
| 12 | Line_0009_0010 | 5 |
| 13 | Line_0009_0014 | 14 |
| 14 | Line_0010_0011 | 10 |
| 15 | Line_0012_0013 | 5 |
| 16 | Line_0013_0014 | 14 |

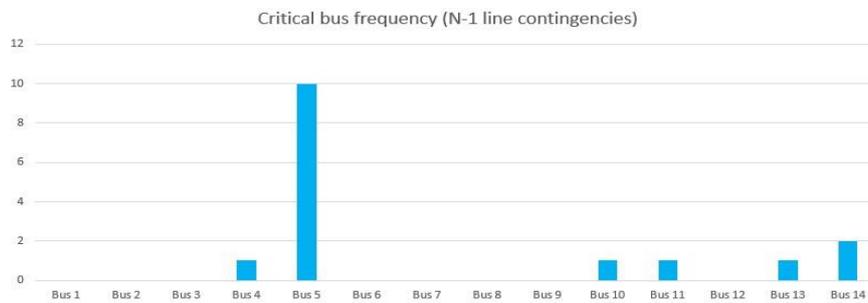

Figure 11. Bar graph showing that Bus 5 is the critical bus for (N-1) line contingencies.

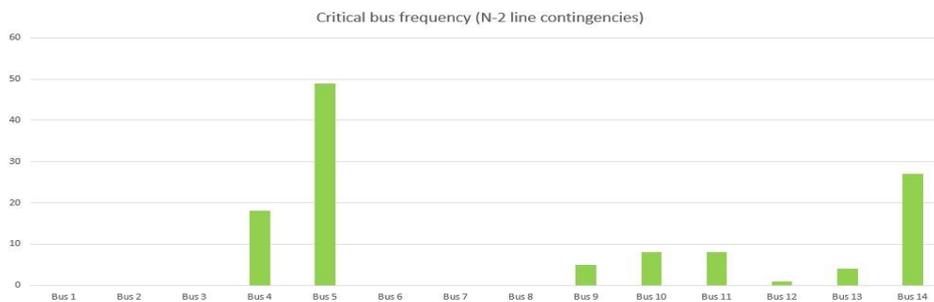

Figure 12. Bar graph showing that Bus 5 is the critical bus for (N-2) line contingencies.

TABLE 3. SUMMARY OF THE P-V ANALYSIS FOR THE IEEE 14-BUS TEST SYSTEM

| Case Type | Critical Bus (without SCs) | Critical Bus (with SCs) |
|---|---|---|
| Base Case | 14 | 5 |
| (N-1) line contingencies | 14 | 5 |
| (N-2) line contingencies | 14 | 5 |

TABLE 4. SUMMARY OF THE P-V ANALYSIS FOR THE IEEE 14-BUS TEST SYSTEM (SCIG INTEGRATION)

| Case Type | Critical Bus (without SCs) | Critical Bus (with SCs) |
|---|---|---|
| Base Case | 3 | 2 |
| (N-1) line contingencies | 5 | 1 |
| (N-2) line contingencies | 3 | 7 |

TABLE 5. SUMMARY OF THE P-V ANALYSIS FOR THE IEEE 14-BUS TEST SYSTEM (DFIG INTEGRATION)

| Case Type | Critical Bus (without SCs) | Critical Bus (with SCs) |
|---|---|---|
| Base Case | 4 | 11 |
| (N-1) line contingencies | 3 | 2 |
| (N-2) line contingencies | 2 | 6 |

TABLE 6. SUMMARY OF THE P-V ANALYSIS FOR THE IEEE 14-BUS TEST SYSTEM (PV INTEGRATION)

| Case Type | Critical Bus (without SCs) | Critical Bus (with SCs) |
|---|---|---|
| Base Case | 8 | 4 |
| (N-1) line contingencies | 2 | 3 |
| (N-2) line contingencies | 7 | 5 |

## 7. CONCLUSION AND FUTURE WORK

The impact of SCs on static voltage stability analysis for the IEEE 14-bus test system was presented in this paper in the presence of both (N-1) and (N-2) line contingencies. The P-V analysis based on the CPF method was used. The critical buses were identified for both cases. In the absence of SCs, it was found that Bus 14 is the critical bus, irrespective of the number of line contingencies (N-1 or N-2). Similarly, it was found that Bus 5 is the critical bus in the presence of SCs, irrespective of the number of line contingencies (N-1 or N-2). A suitable future research direction would be to extend this study to a large-scale real transmission network, which includes generator or transformer contingencies in addition to line contingencies. A comparative analysis for other voltage stability enhancement technologies, such as static VAR compensator (SVC), static synchronous compensator (STATCOM), and unified power flow controller (UPFC), is also another potential area for future research. Artificial intelligence techniques, including deep learning and extreme learning, can be applied to produce a much faster solution.

**Disclosure statement**

No potential conflict of interest was reported by the author.

**Notes on Contributor**

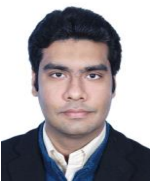

*Umair Shahzad* was born in Faisalabad, Pakistan. He received a B.Sc. Electrical Engineering degree from the University of Engineering and Technology, Lahore, Pakistan, and a M.Sc. Electrical Engineering degree from The University of Nottingham, England, in 2010 and 2012, respectively. He is currently working towards Ph.D. Electrical Engineering from The University of Nebraska-Lincoln, USA. His research interests include power system analysis, power system security assessment, power system stability, and probabilistic methods applied to power systems.